\documentclass[twocolumn,apj]{emulateapj}

\begin{document}

\title{The fast filament eruption leading to the X-flare on March 29, 2014}

\author{Lucia Kleint\altaffilmark{1}, Marina Battaglia\altaffilmark{1}, Kevin Reardon\altaffilmark{2}, Alberto Sainz Dalda\altaffilmark{3},  Peter R. Young\altaffilmark{4} and S\"am Krucker\altaffilmark{1,5}}
\altaffiltext{1}{University of Applied Sciences and Arts Northwestern Switzerland, Bahnhofstrasse 6, 5210 Windisch, Switzerland}
\altaffiltext{2}{National Solar Observatory, Sacramento Peak, P.O. Box 62, Sunspot, NM 88349, USA}
\altaffiltext{3}{Stanford-Lockheed Institute for Space Research, Stanford University, 
HEPL, 466 Via Ortega, Stanford, CA 94305, USA}
\altaffiltext{4}{College of Science, George Mason University, 4400 University Drive, Fairfax, VA 22030}
\altaffiltext{5}{Space Sciences Laboratory, University of California, Berkeley, CA, USA}


\begin{abstract}
We investigate the sequence of events leading to the solar X1 flare SOL2014-03-29T17:48. Because of the unprecedented joint observations of an X-flare with the ground-based Dunn Solar Telescope and the spacecraft IRIS, Hinode, RHESSI, STEREO, and SDO, we can sample many solar layers from the photosphere to the corona.  A filament eruption was observed above a region of previous flux emergence, which possibly led to a change in magnetic field configuration, causing the X-flare. This was concluded from the timing and location of the hard X-ray emission, which started to increase slightly less than a minute after the filament accelerated. The filament showed Doppler velocities of $\sim$2--5 km\,s$^{-1}$ at chromospheric temperatures for at least one hour before the flare occurred, mostly blueshifts, but also redshifts near its footpoints. 15 minutes before the flare, its chromospheric Doppler shifts increased to $\sim$6--10 km\,s$^{-1}$ and plasma heating could be observed, before it lifted off with at least 600 km\,s$^{-1}$, as seen in IRIS data. Compared to previous studies, this acceleration ($\sim$3--5 km\,s$^{-2}$) is very fast, while the velocities are in the common range for coronal mass ejections. An interesting feature was a low-lying twisted second filament near the erupting filament, which did not seem to participate in the eruption. After the flare ribbons started on each of the second filament's sides, it seems to have untangled and vanished during the flare. 
These observations are some of the highest resolution data of an X-class flare to date and reveal some small-scale features yet to be explained.

 \end{abstract}

\section{Introduction}

Solar flares are highly energetic events, which affect a wide range of atmospheric heights. However, no single observatory has the capabilities to measure all layers from the photosphere to the corona for a thorough three-dimensional analysis of solar flares, requiring coordinated joint observations for maximum scientific output. For example, the space-based Solar Dynamics Observatory (SDO) records the full solar disk and thus captures most of the flares, but only with a resolution of $\sim$1\arcsec. While it provides unprecedented context images in multiple wavelengths, a comprehensive study of detailed flare mechanisms requires a higher spatial resolution and different techniques, such as spectroscopy and polarimetry in many different wavelengths. The powerful Interferometric BIdimensional Spectropolarimeter (IBIS) and Facility Infrared Spectropolarimeter (FIRS) at the ground-based Dunn Solar Telescope (DST), which feature spectropolarimetry of photospheric and chromospheric spectral lines at $\sim$0\farcs2, and the recently launched Interface Region Imaging Spectrograph (IRIS), which enables spectroscopy of the solar transition region (TR) at a spatial resolution of $\sim$0\farcs3, are most promising for flare studies. On March 29, 2014, all these observatories, plus Hinode and the Reuven Ramaty High Energy Solar Spectroscopic Imager (RHESSI) recorded the X1 flare SOL2014-03-29T17:48 during our joint observing campaign.

In this paper, we focus on combining data from several instruments to investigate the evolution and possible triggers of the flare, especially a filament eruption. Filaments have been observed to rise before their eruption, accelerating from a slow motion of $<$15 km\,s$^{-1}$  to $>$100~km\,s$^{-1}$  \citep{sterlingmoore2004,sterlingmoore2005}. While the rise phase may last for several hours, a phase of rapid acceleration often occurs within minutes. Together with coronal plasma, such an erupting filament may form a coronal mass ejection (CME), which propagates into interplanetary space at velocities of about 20 to $>$2500~km\,s$^{-1}$  \citep{webbhoward.lrsp}.  

Filaments are made of dense material at chromospheric temperatures, and are probably suspended by magnetic fields, such as at the bottom of a flux rope or on a sheared arcade \citep[e.g.,][]{parenti2014}. A filament eruption is probably triggered by a change in the magnetic configuration of its surroundings. Because the surrounding field currently cannot be directly measured, the filament is used to trace its location. Several studies focused on the filament height evolution (exponential, constant, power-law) to infer velocities and acceleration \citep[e.g.,][]{gilbertetal2000,gallagheretal2003,goffetal2005,williamsetal2005,schrijveretal2008}, trying to explain the eruption with simple scenarios for the magnetic field destabilization, such as breakout, tether-cutting, or different kinds of instabilities. Their different results indicate that the mechanisms responsible for filament acceleration either vary, or that the filament height may not be a good proxy for the processes behind the eruption, as for example the overlying flux configuration may also be important for the eruption. While most of these studies looked at erupting filaments at the limb, where the velocity is determined from the filament height in time, our data allow direct measurements of Doppler velocities, which also show the ranges of the plasma motion. We are therefore not only able to directly determine velocities at chromospheric and TR temperatures, but also to probe the temperature of the filament through the multitude of spectral line observations and RHESSI.

\section{Observations and data reduction}
\label{obs}
\begin{figure} 
   \centering 
   \includegraphics[width=.5\textwidth]{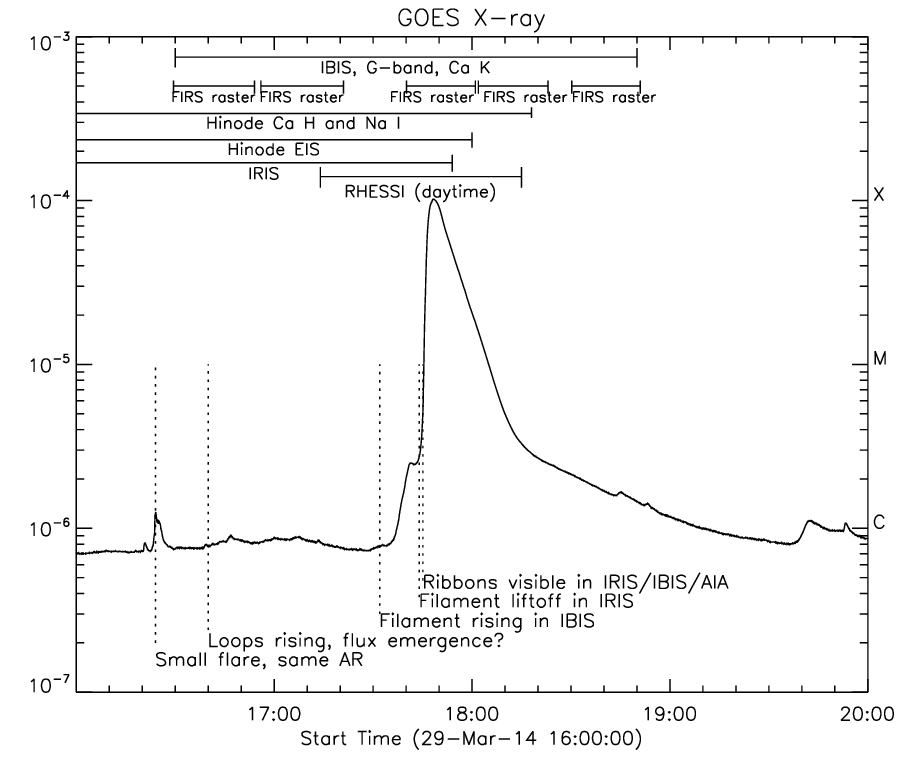}
   \caption{GOES X-ray flux (1--8 \AA) showing the X1 flare on 2014-03-29. The observing times of other instruments are indicated on top, some selected events are labeled on the bottom.}
         \label{goes}
  \end{figure}

We combine data from different instruments to investigate the photosphere, chromosphere, TR, and the corona. Figure~\ref{goes} shows the GOES X-ray flux and the observing periods of each instrument, apart from SDO, which runs continuously.

\subsection{IBIS}
For the high-resolution temporal evolution of the chromosphere we focus on IBIS data. IBIS \citep{cavallini2006, reardoncavallini2008} is a dual Fabry-Perot instrument at the DST with several prefilters available to scan different spectral ranges quasi-simultaneously. Only ground-based instruments, in this case IBIS and FIRS, are currently able to record polarimetric data of the chromosphere. On March 29, 2014, we took data of AR 12017 ($\sim$10.8$^\circ$N, $\sim$33.6$^\circ$W ) from 16:30 until 18:50 UT, capturing all phases of the X1 flare.
\begin{figure*} 
   \centering 
   \includegraphics[width=\textwidth]{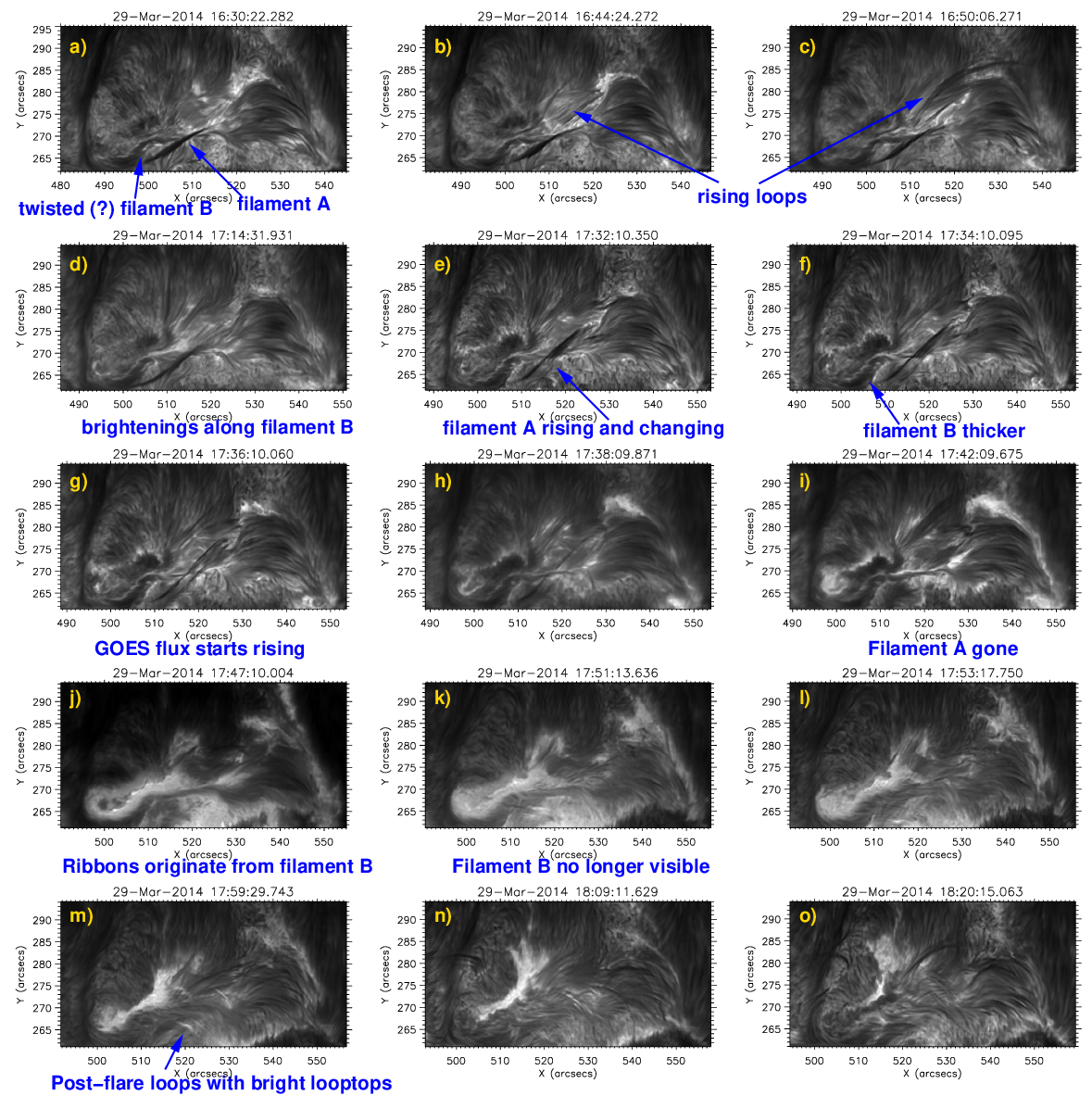}
   \caption{Selected images from IBIS \ion{Ca}{2} 8542 \AA, showing the evolution of the active region before and after the X1 flare. A movie with larger frames is available online.}
         \label{overview}
  \end{figure*}

Our observing sequence included the lines \ion{Ca}{2} 8542~\AA, H-$\alpha$ 6563~\AA, and \ion{Fe}{1}~6302~\AA, with 21, 25, and 23 wavelength points, respectively, which sample the photosphere and the chromosphere. Including the cycle of the six polarization states (for Ca and Fe), the overall cadence ranged from 40 s to 60 s, depending on the number of repetitions of each H-$\alpha$ wavelength. These repetitions were tested as a means to improve the image reconstruction through a Multi-Object-Multi-Frame-Blind Deconvolution. The data reduction steps are described in detail in \citet{kleint2012} and include corrections for dark current, flatfielding, alignment of the different channels, corrections for the instrumental wavelength variation across the field of view (FOV), and for the prefilter transmission profile. Additionally, to obtain the highest spatial resolution possible, a second reduction pipeline was applied to the \ion{Ca}{2}~8542~\AA\ data: for each scan, we took the 30 images with their wavelength closest to the line core, independent of polarization state, and speckle-reconstructed them \citep{woegervdl2008}. These images are used for Fig.~\ref{overview}. IBIS images have a FOV of $\sim$40\arcsec$\times$90\arcsec\ with a plate scale of 0\farcs1/pixel.

A G-band camera for context imaging was run simultaneously and recorded bursts of 20 images within a few seconds once per minute, which were also speckle-reconstructed for a final cadence of one image per minute.

\subsection{IRIS}
IRIS data are provided as Level 2 FITS files, which already include all relevant calibrations \citep{iris2014}. Several images during the flare suffered from overexposure. However, our analysis focuses on the erupting filament, which was not saturated. During this observation, IRIS was performing an 8 step raster, with steps of 2\arcsec, giving a total FOV of 14\arcsec $\times$ 174\arcsec\ for the spectra. Slitjaw images (SJI) were obtained in the wavelengths 1400~\AA\ (at steps 1,5,7), 2796~\AA\ (at steps 2,4,6,8), and 2832~\AA\ (at step 3), sampling the chromosphere and the TR. Each raster step took about 9 s, with the exposure time varying from the nominal 8 s in SJI and near-UV spectra (NUV) due to the automatic exposure control (which however does not influence the duration of one step). The spatial sampling along the IRIS slit is 0\farcs166/pixel. The spectral sampling was 0.025 \AA/pixel.

\begin{figure*} 
   \centering 
   \includegraphics[width=.85\textwidth]{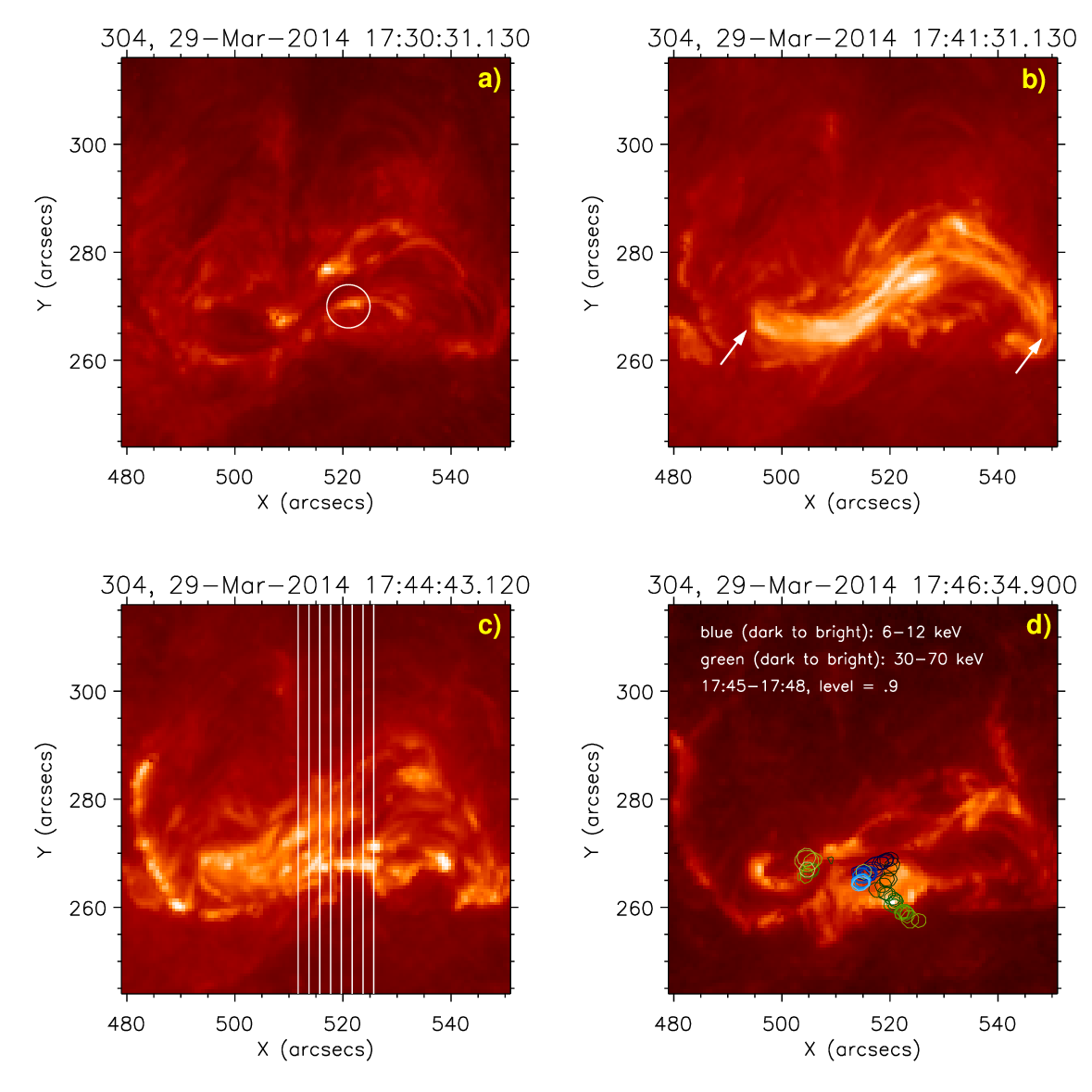}
   \caption{AIA 304 images showing the evolution of the eruption of filament A in four snapshots. a) At 17:30 UT, a bright loop appears near the filament's center (circled). b) After a sudden heating, the filament appears bright at 17:41 UT, and its approximate footpoints are easily visible (arrows). c) At 17:44 UT, the filament erupts. The eight IRIS slit positions are drawn on the image in white. d) The flare ribbons are visible and the image also contains the RHESSI 90 \% contours of the HXR (green) and SXR (blue) emission from 17:45--17:48 UT. Earlier times correspond to darker contours. The northern footpoint is rather stationary, while the southern footpoint first moves parallel to the ribbon, then perpendicularly. A movie is available online.}
         \label{aia304}
  \end{figure*}

\subsection{Hinode and SDO}
The Hinode Narrowband Filtergraph \citep[NFI,][]{tsuneta2008sot} was run in a shutterless mode, recording Stokes I and V in the \ion{Na}{1} 5896 \AA\ line. Each final image of $\sim 60\arcsec \times 80\arcsec$ is a combination of two different exposures, each lasting about six seconds. The cadence of the NFI images is  $\sim32$ s.

The EUV Imaging Spectrometer \citep[EIS,][]{culhaneeis2007} on board the Hinode
spacecraft obtained 104 rasters of the flaring region over the period
14:05 to 18:00 UT, at a cadence of 2 minutes 14 seconds. Eleven steps of
4" were performed with the 2" slit giving a field of view of 42" x 120".
EIS rasters from west-to-east, the opposite direction to IRIS. Eight
spectral windows were downloaded, giving access to the emission lines  \ion{He}{2} 256.28, 
\ion{Mg}{6} 268.99, \ion{Fe}{8} 186.60, \ion{Fe}{12} 186.88, 192.39, \ion{Fe}{16}
262.99, \ion{Fe}{17} 269.42, \ion{Ar}{14} 194.40, \ion{Ca}{15} 181.90, 182.87 and \ion{Fe}{23}
263.76 \AA\ that are formed over the temperature range 0.1 to 15 MK. EIS has a
plate scale of 1"/pixel and a spatial resolution of 3-4". The key EIS
rasters for the filament eruption began at 17:41:46 and 17:44:00 UT, and
the EIS and IRIS slits were approximately at the same location at around
17:42:30 ($x\approx 520$") and 17:44:40 UT ($x\approx 516$").

SDO level 1 data were reduced using {\it aia\_prep} to obtain absolute solar coordinates \citep{lemenetal2012}. All other instruments were aligned with respect to SDO.

\subsection{RHESSI}

RHESSI \citep{linetal2002} soft X-ray (SXR) spectra and images were used to determine the temperature and location of high temperature ($> \sim$8 MK) plasma in the early stages of the filament eruption. Hard X-ray (HXR) images were used to trace the footpoint motion during the flare. 

RHESSI spectra from individual detectors were fitted between 4-11 keV during the rise of the filament, and before both attenuators moved in. The fitted temperatures from the individual detectors give a mean temperature and a confidence interval (standard deviation). For the rise phase, CLEAN \citep{hurfordetal2002} images at 6--12 keV (using grids 3-8, clean\_beam\_width\_factor 2) over a time-interval of 32 s were made to determine the location of the SXR emission (see Figure~\ref{allinstr}). CLEAN images over 12 s intervals at 30-70 keV were made during the HXR peak from 17:45--17:48 UT to trace the footpoint motion. A 0.2$^\circ$ clockwise roll about Sun center was applied to RHESSI images to align them with SDO.
\begin{figure*} 
   \centering 
   \includegraphics[width=\textwidth]{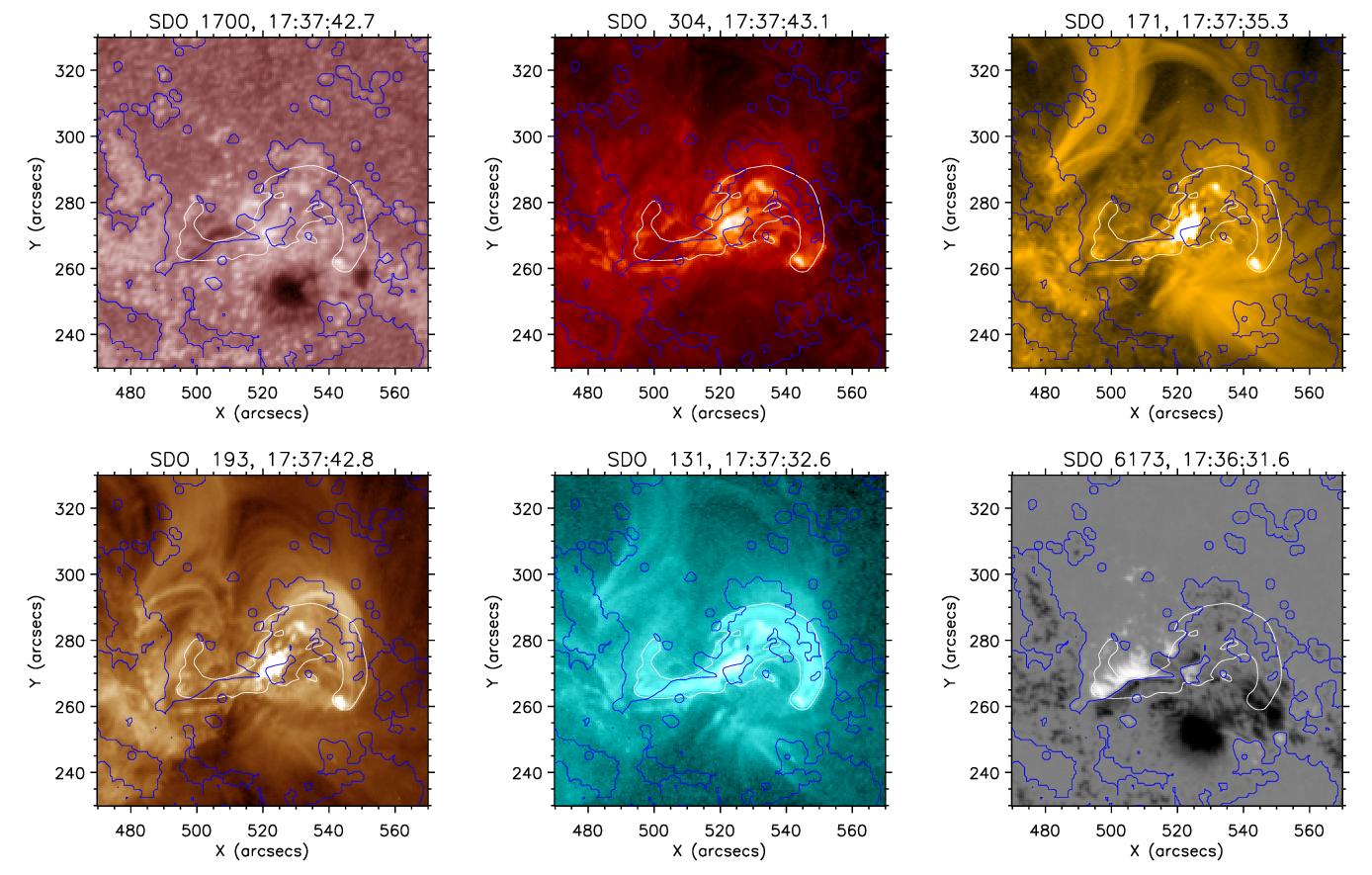}
   \caption{Selected SDO images of the filament after plasma heating occurs around 17:37:35 UT (exact times given in panels). White contours outline the intensity of AIA 131, blue contours show polarity inversion lines and were created by smoothing the magnetogram by 3 pixels and ignoring values of B$_{\rm LOS}$ below 30 G. A movie including all AIA passbands is available online.}
         \label{allaia}
  \end{figure*}

\section{Flare evolution}
 
Figure~\ref{overview} shows the temporal evolution of the chromosphere as seen in \ion{Ca}{2}~8542~\AA. More than one hour before the flare started, a rather stationary filament (labeled as ``filament A'') and what looks like a twisted filament (labeled as ``filament B'') are visible in absorption (panel a). While there are many names for such absorption features in the literature, and one may probably argue that our ``filament'' is also a flux rope, we will stick to these designations throughout our paper for consistency. Our designation ``filament'' also refers to both the absorption feature and later the bright structure seen in IRIS and AIA. Filament B lies above the main polarity inversion line, as determined from Hinode and SDO/HMI magnetograms. Filament A seems to lie slightly further south, although there are patches of mixed polarities below it. The Calcium line is blueshifted by 2--4 km\,s$^{-1}$ along the filament, but at filament A's approximate footpoints (determined from AIA since the filament is not visible fully in Calcium, see Fig.~\ref{aia304}b), there are redshifts of similar size. It is unclear at what time this motion started because it is visible in all our IBIS data since 16:30 UT. 

A small flare occurred around 16:25 UT in the same active region, at a location north of filament A where flux emergence occurs around 16:45 UT. The newly formed loops (Fig.~2b and c) are also visible in the TR as bright features in IRIS SJI images. Around 17:32 UT, filament A starts changing its shape and visibly rises in IBIS 8542 images (2e), before it vanishes at 17:42 UT (2i), probably reaching higher solar layers and higher temperatures. Indeed, it becomes visible in IRIS SJI 1400 at 17:34 UT in emission, and erupts at 17:45 UT. Already before the eruption, the chromosphere becomes bright in several locations (2d, f, g, h), including the two footpoints of filament A. They are visible as the hook-shaped features in the 17:42 UT image (\ref{overview}i), at coordinates [499,268]\arcsec\ and [547,262]\arcsec.

Nearly simultaneously with the eruption, two flare ribbons appeared and seem to originate below filament B, moving away from it perpendicular to its axis (2j and k). Because the ribbons appear close together and between the footpoints of filament A, it is likely that low-lying loops, i.e., loops below the rising filament A, reconnected.

A few minutes later, the seemingly twisted filament B is no longer visible, but seems to be replaced by regular filamentary chromospheric structure (2k and l). It looks as if it simply disentangled during the flare. Also notable is the appearance of post-flare loops with bright loop tops (2l, m, n), while the heating in the chromosphere continues, and parts of the ribbon still expand at 18:09~UT (2n and o).

The four snapshots in Fig.~\ref{aia304} show the evolution of the filament eruption in AIA 304~\AA. The filament eruption is only visible in higher temperatures, such as in AIA passbands and IRIS, and not in the IBIS images. The flare ribbons clearly show a similar structure in AIA 304 and IBIS (Fig.~3d and 2j). IRIS slit positions and RHESSI contours, which will be discussed later, are overplotted on two of the images.

Figure~\ref{allaia} shows an overview of selected SDO/AIA passbands and an HMI magnetogram with the polarity inversion lines drawn as blue contours. White contours outline the filament intensity as seen in SDO/AIA 131. It can be seen that the filament lies above small patches of opposite polarity. In ``cooler'' wavelengths, such as 304 \AA, a bright intensity is visible near the middle of the filament, corresponding to a polarity inversion line just north of a patch of positive (white) polarity, embedded in mostly negative (black) polarity. This is the approximate location of flux emergence that occurred throughout that day (see online HMI movie). In ``hotter'' wavelengths, such as 131 \AA, the whole filament outline is already bright. For the cooler temperatures, this occurred a few minutes later (see online movie).

\begin{figure} 
   \centering 
   \includegraphics[width=.5\textwidth]{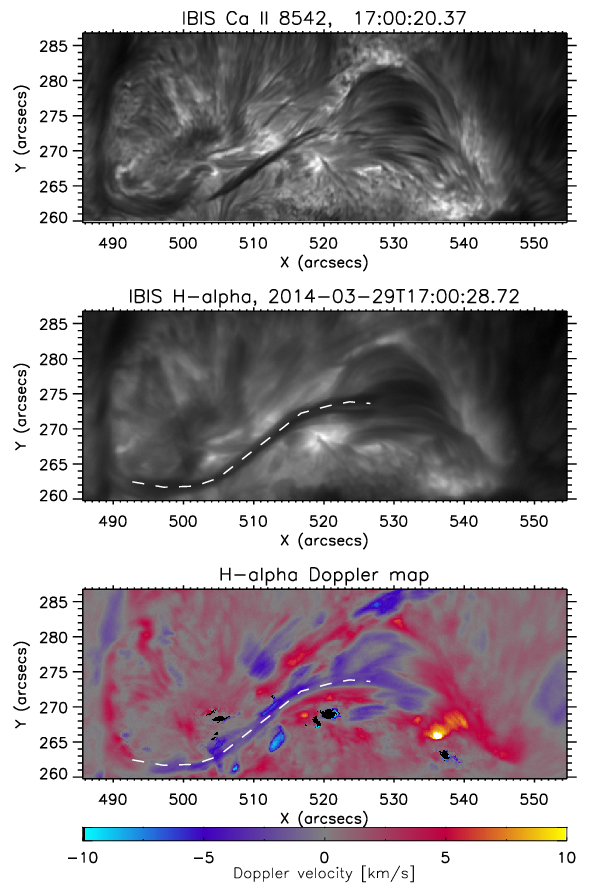}
   \caption{Top: Calcium intensity image where only the central part of filament A (around [510,267]\arcsec) is visible,  which appears darker than filament B (around [502,267]\arcsec). Middle: H-$\alpha$ image taken at the same time showing the whole filament A with its main part marked by a dashed line. Bottom: Doppler map of H-$\alpha$ showing mostly blueshifts along the filament and redshifts at its western footpoint. The velocity scale was clipped at $\pm$10 km/s.}
         \label{hadoppler}
  \end{figure}

\begin{figure*} 
   \centering 
   \includegraphics[width=\textwidth]{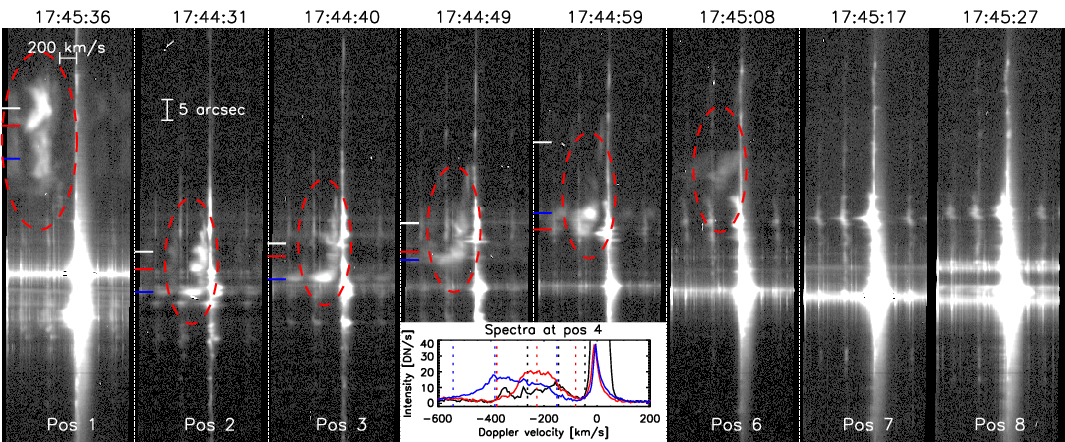}
   \caption{Spectra of the \ion{Si}{4} 1403 \AA\ line and some weaker lines for the 8 raster positions of IRIS, scaled for better contrast. Raster positions 2-8 belong to the same raster and raster position 1 was plotted from the subsequent raster to show the high velocity (blueshift) of the filament (circled in red). In absolute coordinates, the plotted slits span solar y=[239.5,314.2]\arcsec. The small colored lines at the left of most spectra show the color-coded positions where the velocities were determined for Fig.~\ref{vel}. The inset shows the three colored spectra from position 4 with their ranges and centeroids denoted by vertical color-coded lines, to show an example of how the velocities were derived.}
         \label{iris}
  \end{figure*}

\subsection{Filament velocities}

In the following, we investigate filament A in more detail.
The line-of-sight velocity component of filament A can be measured through the Doppler shifts of spectral lines, while the plane-of-sky component can be measured by tracking the apparent position of the filament in image sequences.
We determine the Doppler velocities from IBIS, IRIS, and EIS spectra. These are a lower limit to the actual velocities if the motion is not along the line-of-sight. From the AIA 304 images, we determine the projected motion of the filament during the same time as the IRIS spectra, which only works for a few frames (on-disk) because the filament becomes too faint.

\subsubsection{Doppler velocities}
The IBIS Calcium and H-$\alpha$ intensity spectra show filaments at chromospheric temperatures. The determination of Doppler shifts is not trivial because of the highly variable line shapes, especially for the \ion{Ca}{2} 8542~\AA\ line, which ranged from emission profiles, relatively flat profiles, to absorption profiles, or highly asymmetric shapes. Therefore, we used a machine-learning technique \citep[k-mean clustering;][]{macqueen1967} to classify the different intensity profile shapes, and for each of these classes we determined suitable parameters for a Gaussian fit to find the line center, or omitted them completely from the velocity determination. In this way, we obtained coherent velocity maps that may not be perfect (especially at the borders between emission and absorption profiles, for example), but which showed good fits in the spectral profiles of the filament. They yielded a blueshift of 2--4 km\,s$^{-1}$ , increasing to 5--9 km\,s$^{-1}$  around 17:30 UT for the central part of the filament that was visible in this line. Patches of redshifts were visible near filament A's footpoints, although it cannot be determined if those arise from filament or background motion, because only the central part of the filament was clearly visible in the 8542~\AA\ line (cf. Fig.~\ref{hadoppler}).

For H-$\alpha$, we employed the bisector-technique, calculating 10 bisectors between the minimum and the maximum of the spectral line. From the low bisectors (70-90\% of line depth), we then obtained the velocities. An example from 17:00 UT is shown in Fig.~\ref{hadoppler}. Displayed are (from top to bottom): an intensity image of \ion{Ca}{2} 8542 \AA\ (speckle-reconstructed), an intensity image of H-$\alpha$ (original image), and the corresponding  H-$\alpha$ Doppler map. The main part of the filament is marked with a dashed line. It always showed upflows on the order of $-2$ to $-3$ km\,s$^{-1}$ , also increasing its velocities around 17:30 UT to $-4$ to $-6$ km\,s$^{-1}$  and then to above $-10$ km\,s$^{-1}$. The right (western) footpoint started with downflows, indicating a arch-shaped geometry of filament A and flows along the filament. The H-$\alpha$ images show an intriguing sub-structure of Doppler shifts, which will be analyzed in more detail in a forthcoming publication.

In the He 10830 \AA\ line, which also forms in the chromosphere, the filament has been reported to show an upflow of $-5$ to $-10$ km\,s$^{-1}$  during a scan taken from 16:29 to 16:55 UT \citep{philxflare2014}. 

\begin{figure} 
   \centering 
   \includegraphics[width=.5\textwidth]{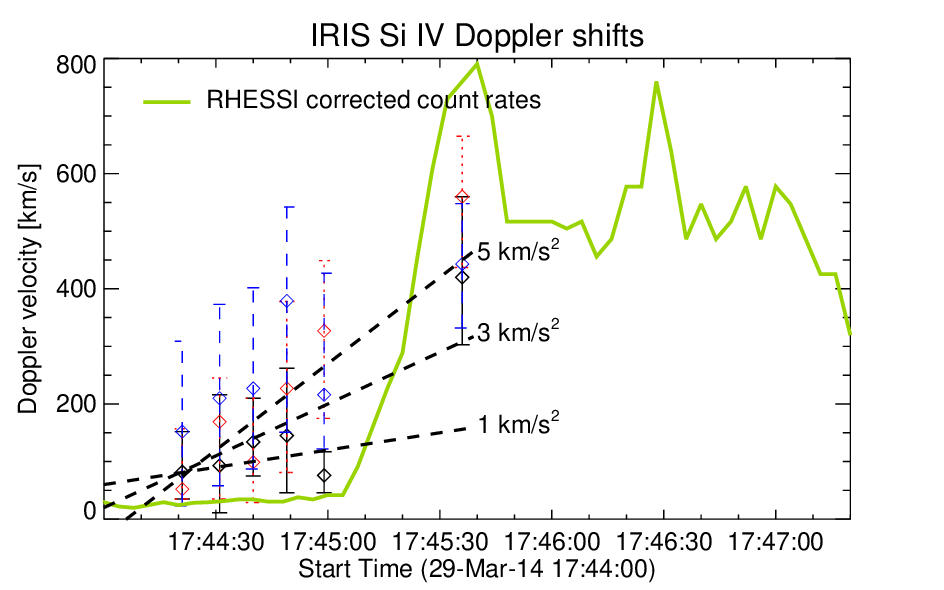}
   \caption{Doppler velocities (upflow), determined from the \ion{Si}{4} 1403 \AA\ line, which forms at about T=10$^{4.8}$ K. Different colors sample three different locations in the filament. The filament lifts off with more than 600 km\,s$^{-1}$. Dashed lines show different constant accelerations for reference. RHESSI corrected count rates for 25-50 keV are overplotted in arbitrary units to show that the filament acceleration started before the impulsive phase of the flare.}
         \label{vel}
  \end{figure}

The IRIS spectra show the liftoff and acceleration of filament A. It is visible in at least 6 different spectra in the lines \ion{Si}{4} 1403 \AA, \ion{C}{2} 1334.5 and 1335.7 \AA, and \ion{Mg}{2} {\em h} and {\em k}, taken during a range of about one minute. Because IRIS scanned in steps of 2\arcsec\ and because each step took about 9 s, our temporal evolution combines different parts of the filament. Figure~\ref{iris} shows 8 raster steps with their observing start times labeled and the signatures of the filament circled in red. The filament was probably already visible in previous spectra, but with lower Doppler velocities and lower intensities in the SJI images, making it impossible to clearly attribute which location along the slit corresponds to the filament. For the spectra shown, we crosschecked the filament location in the 1400 SJI images (where available) and the AIA 304 channel, and then measured the extent of the Doppler shifts of the diffuse emission with respect to the spectral line in a more quiet part of the Sun. The velocities of \ion{Si}{4} were generally higher than those of \ion{C}{2}, usually with a difference of around 10--30 km\,s$^{-1}$  for the centroid of the shifts. However, a direct comparison is very difficult because of the noise levels (\ion{Si}{4} is much brighter than \ion{C}{2} and \ion{Mg}{2}), and the two  \ion{C}{2} lines in the same restricted passband (any velocity above --350 km\,s$^{-1}$  would be out of the FOV in \ion{C}{2} 1334.5, and \ion{C}{2} 1335.7 blueshifts may blend with the former). The diffuse clouds in Fig.~\ref{iris} indicate that large ranges of velocities were present simultaneously. We selected three locations along the slit in each spectrum, aiming for the lower part, the middle and the upper part of the filament, and determined their range of Doppler velocities. The inset shows the color-coded spectra corresponding to raster position 4. The vertical lines indicate how the velocities were determined, which had to be done manually, as the velocity distributions are not Gaussian. The lower velocity limit is given by the blending of the cloud with the spectral line, and it is probable that even lower velocities were present. 
All these Doppler measurements are plotted in Fig.~\ref{vel}, each color denoting one position along the slit, which did not always sample the same part of the filament, because the filament was too faint to trace properly. Similarly colored points should therefore not be understood as a temporal evolution. The diamonds indicate the centroids of the selected velocity distributions. The velocities approximately double in one minute, from $\sim$--200 km\,s$^{-1}$  to over --400 km\,s$^{-1}$. Lines of constant acceleration are plotted for reference and show that the acceleration was approximately between 3 and 5 km\,s$^{-2}$. The green line represents corrected RHESSI count rates at 25-50 keV (units arbitrary), showing that the HXR emission started increasing after the filament was already accelerating, indicating that the flare was a consequence of the instability leading to the filament eruption, and not a cause.

\begin{figure*} 
   \centering 
   \includegraphics[width=\textwidth]{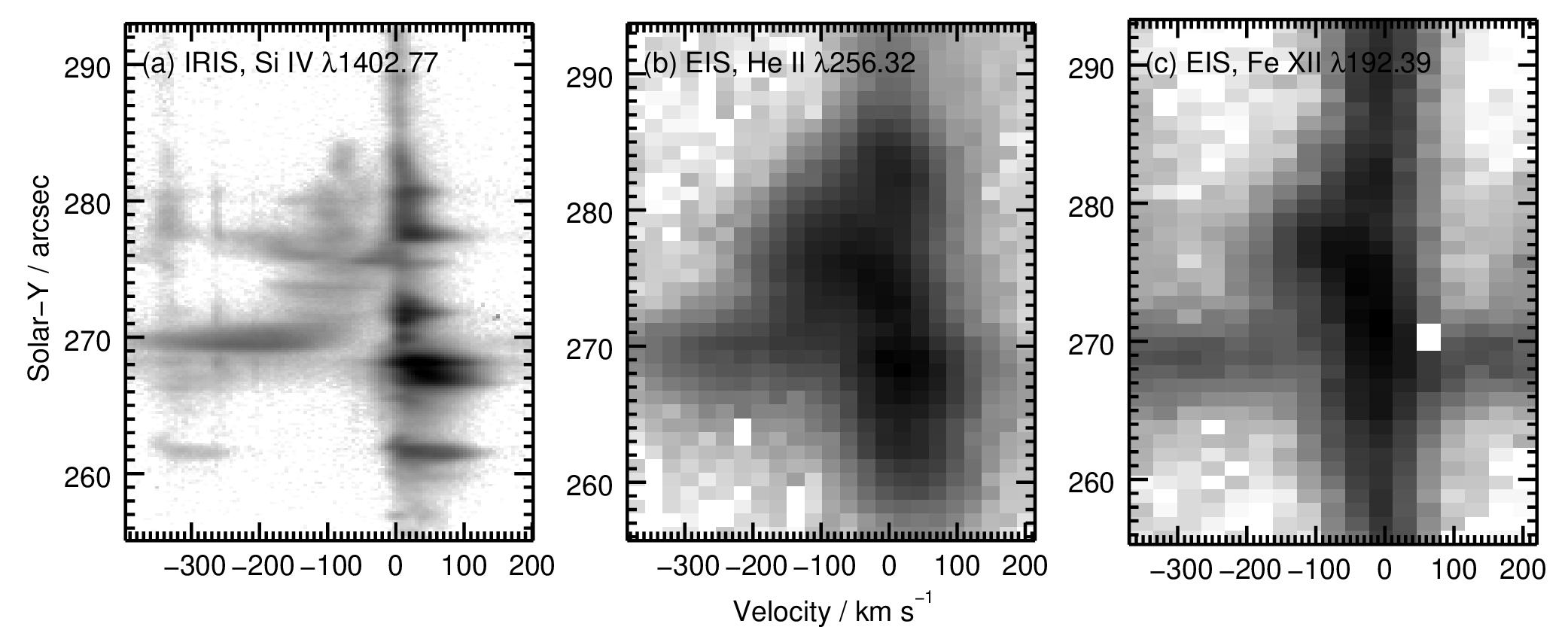}
   \caption{Panel (a) shows an IRIS detector image for Si IV 1402.77 obtained at 17:44:40 UT at about T$\approx$10$^{4.8}$ K. A blue-shifted component is seen at Y=270 with velocity between --380 and --100 km\,s$^{-1}$, and a number of weaker, less blue-shifted features are seen between Y=272 and 284. Panels (b) and (c) show EIS detector images for He II 256.32 (T$\approx$10$^{4.9}$K) and Fe XII 192.39 (T$\approx$10$^{6.1}$K) from 17:44:40 UT. The blue-shifted feature at Y=270 is clearly seen in both lines. A reverse-logarithmic scaling was applied to all images.}
         \label{eis}
  \end{figure*}

EIS confirms the presence of high velocities, which are seen in the rasters from 17:39:32 and 17:41:46 UT. The blueshifts are visible as enhanced blue wings, but the spectral resolution does not allow to clearly separate the shifted component as in the IRIS data. The EIS spectra taken at 17:42:22 (i.e.\,before the IRIS data shown in the Figures) show centroids of the Doppler shifts at --140 to --160 km\,s$^{-1}$  for the spectral lines \ion{He}{2} 256.3, \ion{Fe}{8} 186.6, and \ion{Fe}{12} 186.9 and 192.4, consistent with a slower acceleration at this time. The IRIS and EIS slits crossed around 17:44:40. The EIS He spectra starting at 17:44:36 were therefore taken in a similar location as raster position 3 in Fig.~\ref{iris}, and indeed show a similar structure along the slit, as shown in Fig.~\ref{eis}. The brightest IRIS feature at that time (marked in blue in Fig.~\ref{iris}) has a velocity range of --90 to --400 km\,s$^{-1}$ with a peak intensity at --223 km\,s$^{-1}$ in \ion{Si}{4}. The corresponding EIS feature in \ion{He}{2} 256 has an enhancement from --80 km\,s$^{-1}$  to above --410 km\,s$^{-1}$  (edge of FOV) with no clear centroid. The rest component of the line has the highest intensity and its blue wing is decreasing nearly linearly. The EIS velocities are therefore compatible with the IRIS observations, but not directly comparable. EIS does however allow us to conclude that the emission from the erupting filament is highly multithermal, with strong Doppler shifts being observed from T$\approx$10$^{4.9}$ K (\ion{He}{2}) to T$\approx$10$^{6.1}$ K (\ion{Fe}{12}).

\begin{figure*} 
   \centering 
   \includegraphics[width=\textwidth]{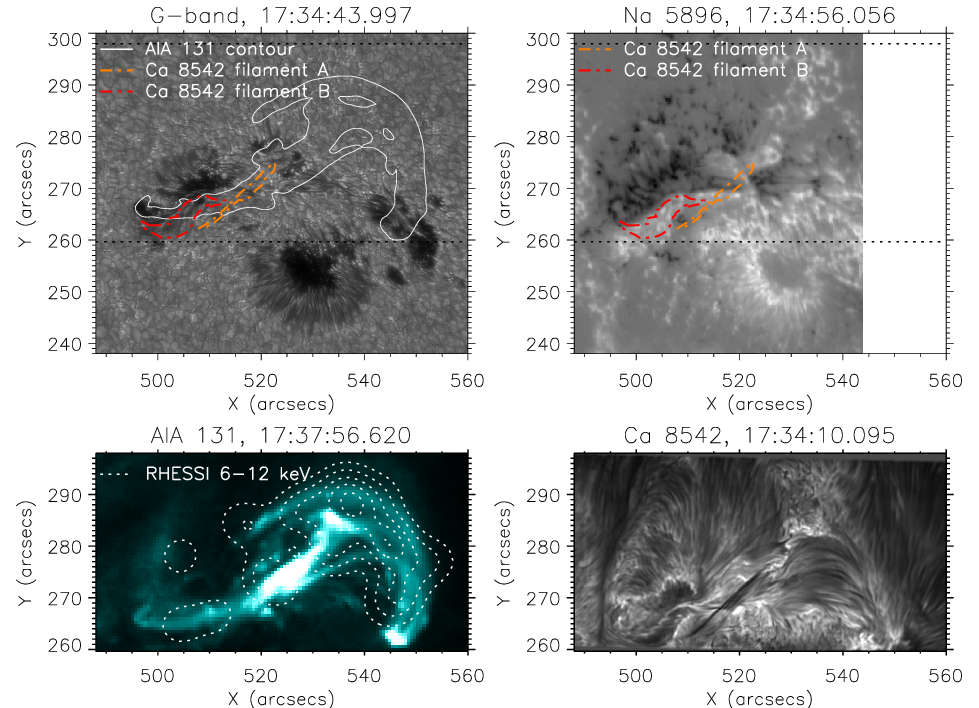}
   \caption{Top left: G-band image from the DST showing AR 12017 in the photosphere. The solid contours show the coronal structure of the filament in AIA 131, which is also shown as an image on the bottom left. The dotted contours show the twisted filament B and filament A from the IBIS Ca 8542 image (bottom right). Top right: Hinode \ion{Na}{1} 5896 Stokes $V$ image (note that the colors are opposite to the HMI magnetogram). The twisted  filament B lies  exactly along the main neutral line, while filament A lies slightly south of it, but crosses locations of smaller opposite polarities. Bottom left: RHESSI contours (dotted) trace the filament (background image), indicating heating.}
         \label{allinstr}
  \end{figure*}

\subsubsection{Projected motion of the filament}

We determined the apparent motion of filament A between 17:44:31 UT and 17:45:19 UT from AIA 304 images. For each of the 5 frames, we manually determined the upper limit of the filament arc at solar coordinates x=[515,520,525,530,535]\arcsec. We then converted the arcseconds into km (using 1\arcsec=725 km), and calculated the velocities. They ranged from 130--230 km\,s$^{-1}$  at 17:44:37 to 340--700 km\,s$^{-1}$  at 17:45:13 UT with the center of the filament moving faster at later times. These velocities are comparable to those of the IRIS \ion{Si}{4} Doppler shift, and possibly even larger than the Doppler velocities after 17:45 UT. The error bars for this method are on scales of $\sim$0\farcs5, which may result in an error of $\pm 30 $ km\,s$^{-1}$ . Because it is unlikely that the filament is moving perfectly parallel or perpendicular to our line-of-sight, then all velocities reported here are lower limits for the true velocity vector.

\subsection{Multi-instrument analysis}

The apparent footpoints of filament A were in the eastern and western sunspots of the AR, shown in the top left panel of Figure~\ref{allinstr}. SDO/HMI magnetograms show that flux emergence and cancellation occurred continuously between these footpoints for more than a day, while the sunspots and thus footpoints drifted further apart from each other, probably twisting and stressing the connecting coronal loops. SDO/AIA images, especially AIA 304, show apparent flows into the footpoints, which periodically become brighter. These flows are already present more than one hour before the flare.

At 17:30 UT, several AIA wavelengths show bright emission with a size of approximately 3\arcsec\ appearing right next to the middle of the filament (circled in Fig.~\ref{aia304})  and above a region of opposite polarities. It looks as if the emission is coming from the top of a loop connecting the western filament footpoint with a region of flux emergence/cancellation around coordinates [520,270], best seen in AIA 304. This is about the time when filament A starts changing, possibly due to being above that hot loop.

The SXR emission at 6--12 keV began to rise at $\sim$17:34~UT, peaking at 17:40, then decreased again until 17:43 when it started to rise at the onset of the flare. The RHESSI spectral fits suggest a maximum temperature of 23 $\pm$ 3 MK during this phase, decreasing to 15 $\pm$ 1.5 MK by 17:43 UT. The bottom left panel of Figure~\ref{allinstr} shows the RHESSI contours (15, 30, 50, 70, 90 \%) overlaid on AIA 131, indicating that the initial heating is occurring  along, or possibly below the filament shape. This is also supported by Fig.~\ref{allaia}, which shows part of the filament still in absorption in AIA 304 with a bright emission below its center, while AIA 131 shows hot plasma along the whole filament shape. Note that the actual flare did not occur at the location of the initial heating but closer to the eastern part of the AR. AIA images reveal that the filament started becoming brighter near its center first, rapidly followed by its right (western) footpoint, as seen in Figure~\ref{allinstr}. The filament center brightening occurs right next to the bright loop from 17:30 UT, making it plausible that reconnection of newly emerging loops with the pre-existing, filament-carrying field caused the strong increase in temperature. Around 17:40 UT, most AIA wavelengths show the intensity of the filament to start spreading from its center to both footpoints, until it is fully lit up at 17:41 UT.  Because the filament fades in chromospheric IBIS images gradually from 17:32 UT to 17:42 UT, and appears in IRIS 1400 SJI at that time, it is possible that at least part of the actual filament material is being heated.

The red and orange dotted contours in Figure~\ref{allinstr} trace the  seemingly twisted filament B and filament A, respectively, and were obtained from the bottom right image showing a snapshot from IBIS \ion{Ca}{2} 8542 \AA. 

\begin{figure} 
   \centering 
   \includegraphics[width=.5\textwidth]{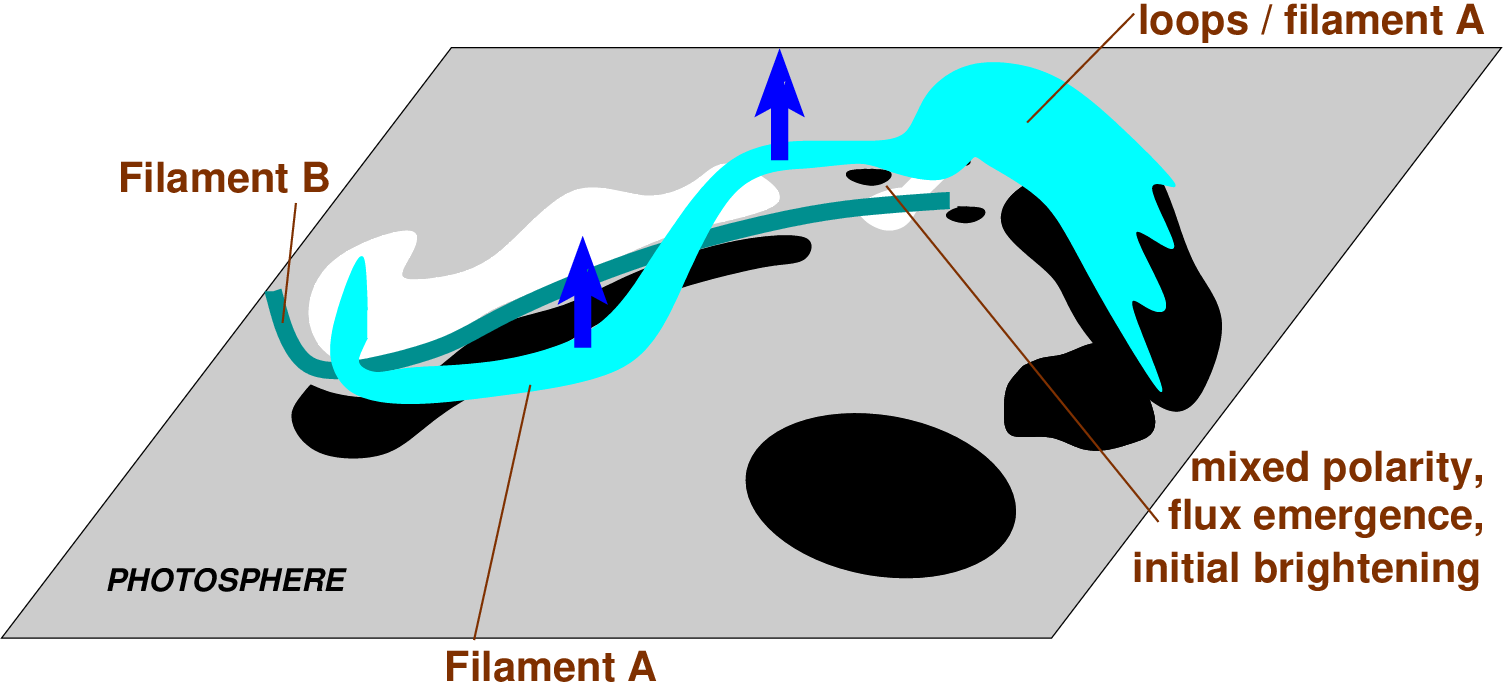}
   \caption{Simplified picture of both filaments (labeled A and B), above the different photospheric polarities (black and white). Filament B may extend further North on the left edge of the image. The blue arrows indicate the eruption direction of filament A.}
         \label{sketch}
  \end{figure}

 A simple sketch summarizing our observations is shown in Fig.~\ref{sketch}. Both filaments, A and B are drawn in color and the eruption direction of filament A is indicated with blue arrows. The black and white features indicate polarities from a photospheric magnetogram, with the large sunspot on the bottom right.

\section{Discussion}

Many of our observed features fit very well into established models of filament eruptions \citep[see reviews by][]{parenti2014,schmiederetal2013,chen2011}. We observe the initial slow rise phase, a sudden acceleration to common CME velocities, and the flare starting. 

\subsection{Filament Destabilization and Flare}
We believe that the filament destabilization is closely related with the flux emergence and cancellation occurring around the pores in the middle of the active region. The emergence resulted in constantly rising and changing loops, one of which may have had the right configuration to start the reconnection and eruption process. The loop system that became bright at 17:30 UT and whose one footpoint was near the emergence/cancellation region was the only visible major change at that time. Because this coincides with the time that the filament started rising, it is possible that the two are related.

The filament itself and its surrounding loops were undergoing a thermal instability for several hours before the flare, as apparent plasma motions are seen into its footpoints  in AIA images. As the intensity near the filament started increasing, hot plasma may have filled the filament channel. By filament channel we mean the volume where the filament forms, which may contain a flux rope \citep[e.g.,][]{tandberghanssen1995}.
While part of the original cooler filament seems to have remained (absorption seen in AIA 304 until $\sim$17:40 UT), some overall heating must have occurred, because it started disappearing in (chromospheric) IBIS data after $\sim$17:35 UT and appearing in IRIS data.

These observations are compatible with many previous filament studies, which, for example, focused on plasma falling from the prominence or heating prior to eruption \citep[e.g.,][]{tandberghanssen1995,Cirigliano2004}, magnetic flux cancellation prior to the destabilization \citep{Huangetal2014}, or SXR flux enhancements  \citep{roytang1975}.
 
The erupting filament probably triggered other reconnection events, very likely located below it because the flare ribbons started very close to each other indicating small and low loops connecting them. Such a reconnection event may have been the cause of the X1 flare. Such a scenario agrees with the post-flare loops seen in \ion{Ca}{2} 8542 (and several AIA passbands) with their bright loop tops and the location of the looptop SXR emission at 6-12 keV from RHESSI during and after the flare. The RHESSI HXR motion of the southern footpoint was parallel to the neutral line at first for about one minute and then perpendicular to it, while the northern footpoint remained rather stationary in the penumbra of the eastern sunspot (see Fig.~\ref{aia304}). Parallel and perpendicular HXR motion may be an indication for asymmetric and symmetric filament eruptions, respectively \citep{liuetal2010}. During this flare, AIA images show that the eruption was not perfectly symmetric, which may explain the parallel motion in the beginning.

\subsection{Twisted Fliament Vanishing}
There are several open questions, one of which is what the role of the  seemingly twisted filament B was. The high-resolution IBIS images show that its structure differs from filament A and it appears to have several locations where intensity variations can be traced nearly perpendicularly to its axis, which is our main indication of twist. In the photosphere, it corresponds to an oddly-shaped penumbra as seen in Fig.~\ref{allinstr}. 
However, this penumbra does not change significantly during the flare, unlike
some of the penumbrae near the central pores, which disappear during the flare.
It is possible that the twisted filament is only a chromospheric phenomenon. Once the filament becomes invisible in IBIS (see 17:42~UT in Fig.~\ref{overview}), it is clearly visible that it extends quite far to the east (coordinates [525,270] in that image). In fact, it looks like a  low-lying filament, unaffected by the first eruption. The reason for its disappearance around 17:50 UT remains unclear and one can only speculate that the rearrangement of loops must also have rearranged it, but without any further reconnection, which would have appeared bright due to heating.


\subsection{High Acceleration:  $\sim$3--5 km\,s$^{-2}$}
The rise velocity is one of the most intriguing results of our study. It is probably the first direct measurement of Doppler velocities of an erupting filament at TR temperatures. Previous studies mostly focused on height-time measurements of erupting prominences above the solar limb \citep{gilbertetal2000, gallagheretal2003, goffetal2005}. Because of the on-disk location of our filament, and the position of the STEREO satellites, which only saw the CME, we cannot determine the absolute height of the filament. Filaments are known to be made of cooler material, embedded in the $\sim$1 MK corona, giving a vast range of possible heights. It is worthwhile to note that during the eruption, the plasma must have been multi-thermal, ranging at least from 10$^4$ (\ion{Mg}{2} formation temperature) to 2$\times 10^7$ K (measured by RHESSI). It is likely that we did not see the filament at chromospheric and TR heights  (for which we assume 500-2000~km and 2000-2500 km), also because the formerly measured eruption velocities at those heights are below 50 km\,s$^{-1}$. To investigate if our filament eruption was especially fast or slow, we can compare our velocities with previous measurements during the onset of the impulsive phase. \citet{kahleretal1988} studied four (on-disk) filaments, and obtained speeds of less than 100 km\,s$^{-1}$  at the start of the impulsive phase, from height-time measurements. Apart from one flare, incidentally also an X-flare and their strongest example, their acceleration values (0.2, 0.3, 1.5, 4.0 km\,s$^{-2}$) were several factors lower than ours ($\sim$3--5 km\,s$^{-2}$). An X-flare filament eruption analyzed by \citet{gallagheretal2003}, yielded higher velocities (above 1000 km\,s$^{-1}$), but lower accelerations ($<$1.5~km\,s$^{-2}$). We therefore conclude that our analyzed filament eruption was relatively fast compared to previous filament eruption speeds during impulsive phases, but well within common CME velocities. The final eruption velocity may have been even higher after the filament was out of the field-of-view of IRIS and IBIS. It should be noted that the acceleration started before the impulsive phase, defined as the onset of the 25-50 keV HXR signal, indicating that the flare was a consequence, and not a cause of the eruption.

We purposely avoid comparing our acceleration with CME observations, because filaments and CME's may be separated by several dozen arcsec, seen nicely for example in Fig.~1 of \citet{temmeretal2008}. While some CME's are linked to filament eruptions, the exact interaction is unclear. Investigations focusing on the timing between HXR and CME acceleration showed that the maximum acceleration is generally reached close to the HXR maximum with error bars of a few minutes due to the lower cadence observations \citep{temmeretal2010}.  According to the automated CME catalog ``Cactus'' \citep[sidc.oma.be/cactus/,][]{robbrechtetal2009}, velocities of $\sim$330-820 km\,s$^{-1}$ were measured during the CME that followed our observed filament eruption. 

Could our Doppler velocities be influenced by plasma flows along the filament? In the chromospheric images (Calcium and H-$\alpha$), the velocities do vary along the filament, including occasional redshifts, but not at large values (few km\,s$^{-1}$).  Flow speeds of several tens of km\,s$^{-1}$ have been observed along prominences in H-$\alpha$ \citep[][and references therein]{tandberghanssen1995}, which are compatible to our observations if we assume a nearly horizontal filament with respect to the solar surface. It is therefore possible that the observed Doppler shifts can be attributed to flows along the filament, rather than the filament motion itself, at least before 17:30 UT. For the IRIS images, we only obtain the velocities at the slit location, and not in the outer parts of the filament. Therefore, it cannot be determined if redshifts/downflows were present in other parts of the filament. However,  in at least a width of 10\arcsec\ (6 slit positions), the filament only shows blueshifts/upflows and since its projected position is changing after $\sim$17:30 UT, at least part of the measured Doppler velocities are very likely true motions. The projected speed of motion of the filament in AIA 304 agrees with our Doppler velocities, indicating that this eruption was indeed especially fast.



\section{Conclusions}\label{results}

We showed the sequence of events leading to the X1 flare on 2014-03-29 by analyzing data from the photosphere to the corona. A  field destabilization visible as filament eruption was the probable cause for this flare, given the timing of the eruption and the subsequent appearance of HXR emission below the filament. Apparent plasma motion into the filament's footpoints hours before the event indicate a thermal instability. Small Doppler shifts along the filament were observed for more than one hour before the flare, indicating flows along the filament and/or a slow rise. After a rising loop system above a flux emergence site possibly reconnected with or near the filament-carrying field, a sudden brightening and heating of the whole filament channel followed before it erupted. 

This is the first X-flare observed by IRIS and the \ion{Si}{4} spectra clearly show Doppler shifts of --100 to --600~km\,s$^{-1}$  during the filament eruption. The wide ranges of Doppler velocities at each raster step indicate that parts of the filament move at different speeds. Projected velocities from the filament motion in AIA 304 are similar to the derived Doppler shifts. It is therefore possible that the actual velocity was even higher. Compared to previous studies, this eruption seems to have been especially fast already early in the impulsive phase and had a comparably high acceleration ($\sim$3--5 km\,s$^{-2}$). Our final measured velocity of --600 km\,s$^{-1}$  agrees with average velocities of CMEs, but we cannot exclude a further acceleration or deceleration outside of our FOV.

It is also very likely the first high-resolution (0\farcs2) observation of an X-flare by a ground-based observatory. The \ion{Ca}{2} 8542 \AA\ images show a fine-scale structure of the chromosphere with many changing features prior and during the flare. An unexplained feature is what looks like a  twisted filament (B) lying low in the chromosphere, below the erupting filament (A). While filament B does not actively participate in the flare, it is the origin of the flare ribbons, and it vanishes minutes after the flare by seemingly untangling.

This is an amazingly rich dataset and ideally suited to study many aspects of a strong flare. Future studies could focus on determining the magnetic field configurations and changes, as there was a significant amount of flux emergence and cancellation. Simulations of very fast filament eruptions and
a quantitative analysis of the energy deposition could shed light on their influence on the magnitude of the observed flare. The high-resolution ground-based images may help to investigate if possibly only small changes in the photosphere and in the chromosphere may contribute to triggering powerful solar flares.


\acknowledgments
This work was supported by a Marie Curie Fellowship and the NASA grant NNX13AI63G. We are very grateful to the observers Doug Gilliam and Mike Bradford at the Dunn Solar Telescope. We would like to acknowledge the fruitful discussions during an ISSI meeting about high-resolution chromospheric observations. SK and MB acknowledge the support by the Swiss National Science Foundation (200021-140308), by the European Commission through HESPE (FP7-SPACE-2010-263086), and through NASA contract NAS 5-98033 for RHESSI. PRY acknowledges funding from NASA grant NNX13AE06G. We thank the anonymous referee for useful comments. We acknowledge the work of all the instrument teams of IBIS, IRIS, SDO, Hinode, and RHESSI and their participating institutions.

\bibliographystyle{apj}
\bibliography{journals,ibisflare}



\end{document}